\documentclass[prb,twocolumn,showpacs,preprintnumbers,amsmath,amssymb]{revtex4}
\usepackage{dcolumn}
\usepackage{bm}
\usepackage{graphicx}
\usepackage{bm}
\begin{document}
\pacs{75.47.Lx, 75.60.Ch, 75.40.-s}
\title{Selective substitution in orbital domains of a low doped manganite : an investigation from Griffiths phenomenon and  modification of glassy features} 

\author{K. Mukherjee and A. Banerjee}
\affiliation{UGC-DAE Consortium for Scientific Research (CSR)\\University Campus, Khandwa Road\\
Indore-452017, M.P, India.}
\date{\today}
\begin{abstract}
An effort is made to study the contrast in magnetic behavior resulting from minimal disorder introduced by substitution of  2.5\% Ga or Al in Mn-site of La${_{0.9}}$Sr$_{0.1}$MnO${_3}$.  It is considered that Ga or Al selectively creates disorder within the orbital domains or on its walls, causing enhancement of Griffiths phase (GP) singularity for the former and disappearance of it in the later case. It is shown that Ga replaces Mn$^{3+}$ which is considered to be concentrated within the domains, whereas Al replaces Mn$^{4+}$ which is segregated on the hole-rich walls, without causing any significant effect on structure or ferromagnetic transition temperatures. Thus, it is presumed that the effect of disorder created by Ga extend across the bulk of the domain having correlation over similar length-scale resulting in enhancement of GP phenomenon. On the contrary, effect of disorder created by Al remains restricted to the walls resulting in the modification of the dynamics arising from the domain walls and suppresses the GP. Moreover contrasting features are observed in the low temperature region of the compounds; a re-entrant spin glass like behavior is observed in the Ga doped sample, while the observed characteristics for the Al doped sample is ascribed only to modified domain wall dynamics with the absence of any glassy phase. Distinctive features in third order susceptibility measurements reveals that the magnetic ground state of the entire series comprises of orbital domain states. These observations bring out the role of the nature of disorder on GP phenomenon and also reconfirms the character of self-organization in low-doped manganites.
\end{abstract}
\maketitle
\section {Introduction}
The self-organized regimes of low doped manganites forms one of the interesting area to study the correlated behavior of spin, charge and orbitals.\cite{dag} Another appealing issue in condensed matter is the emerging similarities in the nature of self-organized structures in different transition metal oxides. Current investigation from NMR, neutron scattering and Raman spectroscopy have revealed an orbital ordered state comprising of ferromagnetic (FM) insulating domains separated by ferromagnetic metallic walls in a single crystal of La${_{0.8}}$Sr$_{0.2}$MnO${_3}$ \cite{pap} and in LaMnO${_{3+\delta}}$.\cite{choi} Distinctive signatures of such self-organized regimes, in the form of orbital domain (OD) state having hole rich (metallic) walls separating the hole deficient (insulating) regions, is also reflected in the bulk magnetic measurements of La${_{0.9}}$Sr$_{0.1}$MnO${_3}$.\cite{muk} Similar type of phase separation in the form of stripes where the holes are collected in domain walls separating antiferromagnetic anti-phase domains is observed in microscopic studies on cuprates.\cite{tran} Studies on nickelates have also revealed an ordering consisting of charged domain walls that forms antiphase boundaries between antiferromagnetic domains due to microscopic segregation of doped holes.\cite{tra} One significant point in all these respect is that the nature of this regime is different from the conventional FM domains which arises purely due to minimization of magnetostatic energy. Another important observation is that these self-organized arrangements seems to be destroyed at higher doping levels as is observed in cuprates where the macroscopic phase separation between superconducting and non superconducting phase vanishes with increase in doping.\cite{chou} Similarly, the OD state in La${_{0.9}}$Sr$_{0.1}$MnO${_3}$ is destroyed with the increase in self-doping, however, it changes the crystallographic structure as well.\cite{muk} Hence it would be interesting to probe the modification in the OD state by suitable quenched disorder within the same crystallographic structure.
 
Effective way to introduce disorder in manganite is through the Mn-site substitution. Earlier studies on various Mn-site substitution have revealed that the induced physical properties depend both on the electronic configuration of the dopants as well as on the change in structure arising from ionic size mismatch.\cite{heb} Hence, incorporation of simple disorder without adding magnetic interactions and lattice distortion, is a non-trivial task because of complexities in manganites. Moreover, another manifestation of disorder is the recent observation of Griffiths phase (GP) like singularities.\cite{chan, sal, sal1,  dei} Such GP regimes is defined in between the completely disordered high temperature paramagnetic (PM) region and long range magnetic ordering temperature. The new temperature scale is called `Griffiths temperature' (T$_G$), where a clean system would have its transition in the absence of disorder.\cite{gri} Significantly, theoretical studies have shown the importance of correlated disorder in generating and enhancing the GP like singularity \cite{voj,san}, particularly around low doping.\cite{bou}

The OD of La${_{0.9}}$Sr$_{0.1}$MnO${_3}$ is supposed to be made of Mn$^{3+}$ rich domains separated by Mn$^{4+}$ rich walls. In this study, the Mn-site substitution are selected in such a way that they will replace either of the Mn ions without changing the structure. Substitution of Al or Ga at the Mn-site is expected to replace Mn$^{4+}$ or Mn$^{3+}$ respectively because of matching of ionic radii.\cite{sunil} Furthermore, for both these cases the resultant spin and orbital moment is zero, stressing the fact that no additional interaction in terms of spin orbit coupling arises due to these substitutions. Though reports indicate that non-magnetic substitutions give rise to the same effect in the resulting compounds,\cite{heb} we observe contrasting magnetic behavior in Al or Ga substituted (2.5\% each) La${_{0.9}}$Sr$_{0.1}$MnO${_3}$ without change in structure. In the parent compound GP phenomenon is observed and this feature is not unusual as it has also been detected in the corresponding single crystal.\cite{dei} Here we show that Ga substitution enhances the GP singularity, whereas it disappears in the Al substituted sample. This is intriguing since these substitutions have insignificant effect on the PM to FM transition temperature (T$_C$). This difference is attributed to the nature of disorder created by Ga and Al substitutions. It will be argued that the disorder created by substitution of Ga has correlation within the Mn$^{3+}$ rich domains since it predominantly replaces Mn$^{3+}$.  On the contrary, correlation of the disorder does not extend to the bulk when Al is substituted since it selectively replaces Mn$^{4+}$ in the hole rich domain walls. Moreover, the low temperature region of the Ga substituted compound show reentrant spin glass like features. In contrast, glassy characteristics are absent in the low temperature state of the Al substituted compound. These contrasting magnetic state at low temperature in the substituted compounds also arise due to the selective substitution at the domains and at their walls. All these results, also points to the intrinsic nature of the OD state where hole rich walls separates the hole poor regions and disorder can be selectively created by proper substitution.

\section {Sample preparation and Characterization}
The La${_{0.9}}$Sr$_{0.1}$MnO${_3}$ sample (parent) is the same as used in ref [4]. The doped samples La${_{0.9}}$Sr$_{0.1}$Mn$_{0.975}$Ga$_{0.025}$O${_3}$ (2.5\% Ga) and La${_{0.9}}$Sr$_{0.1}$Mn$_{0.975}$Al$_{0.025}$O${_3}$ (2.5\% Al) are prepared under the similar conditions. X-Ray diffraction studies on the samples indicates that the entire series is crystallographically single-phase and the pattern collected is analysed by the Rietveld profile refinement.\cite{you} Estimation of Mn$^{3+}$/Mn$^{4+}$ is done by iodometric redox titration using sodium thiosulphate and potassium iodide. Homemade setups  are used to measure ac susceptibility\cite{ash} and dc magnetization\cite{krs}, along with a commercial 14T Vibrating Sample Magnetometer (VSM) from Quantum design.  

All the samples crystallize in orthorhombic structure (Pbnm). Table 1 summarizes the relevant structural parameters obtained by
fitting the powder XRD data by rietveld refinement which shows the changes in the lattice parameters a, b, and c is respectively less than 0.04\%, 0.08\% and 0.02\%. This illustrates that Al and Ga substitutions introduces insignificant changes in the structure. The percentage of Mn$^{3+}$ and Mn$^{4+}$ found from iodometry shows a preferential replacement of Mn$^{4+}$ with Al$^{3+}$ and Mn$^{3+}$ with Ga$^{3+}$. Such preferential replacement of Mn$^{4+}$ and Mn$^{3+}$ by Al$^{3+}$ and Ga$^{3+}$, specially around lower percentage of Mn-site substitution has also been observed in other sample series like Pr$_{0.5}$Ca$_{0.5}$Mn$_{1-x}$Al$_{x}$O$_3$ \cite{sunil} and Pr$_{0.5}$Sr$_{0.5}$Mn$_{1-x}$Ga$_x$O$_3$ \cite{pra}.

\begin{table}
\caption{\label{tab:table 1}Structural and fitting parameters determined from rietveld profile refinement of powder XRD pattern for the sample series. S is goodness of fit. The percentage of Mn ions was determined by redox iodometric titration.}
\begin{ruledtabular}
\begin{tabular}{ccccc}
Samples  &Parent & 2.5\% Al  &2.5\% Ga\\
\hline
a(A${^O}$) &5.5352(2) &5.5360(2) &5.5378(2)  \\ 
\hline
b(A${^O}$) &5.5166(2) &5.5146(2) &5.5215(2)  \\ 
\hline
c(A${^O}$) &7.7924(2) &7.7930(3) &7.7904(3) \\
\hline
V(A${^O}$) &237.94 &237.91 &238.21 \\
\hline
S &1.25 &1.27 &1.19  \\
\hline
Mn${^4}$$^+$\% &11.3 &9.6 &10.7 \\
\hline
Mn${^3}$$^+$\% &88.7 &87.9 &86.8  \\
\end{tabular}
\end{ruledtabular}
\end{table} 

\maketitle\section{Results and discussions} 
\subsection{Distinctive magnetic features arising out of selective substitution at domains and walls  by Ga and Al}
  
Figure 1 shows the temperature dependence of the real part of ac susceptibility ($\chi$$_1$$^R$). All samples undergo a paramagnetic to ferromagnetic transition, around the same temperature, which is characterized by a sharp change in susceptibility near the T$_C$. The parent and the Ga doped sample show another fall in $\chi$$_1$$^R$ at low temperature which is absent for Al doped sample where a very sharp monotonic decrease is observed in $\chi$$_1$$^R$ below T$_C$. This shows the distinct behavior of the series having minimal non-magnetic disorder and similar structural parameters. Even though the nature of $\chi$$_1$$^R$ curve of both the parent and the 2.5\% Ga sample is qualitatively similar, the nature of the low temperature magnetic states are different. While the low temperature magnetic state of the parent shows glassy behavior arising from orbital domains, 2.5\% Ga shows a re-entrant spin glass like behavior. On the contrary, the 2.5\% Al compound shows absence of any glassy features in the low temperature region. The details of the low temperature behavior of the substituted compounds will be described in the later part of the manuscript. For the parent compound the self-organized regimes are of the form of orbital domains .\cite{muk} Generally for a phase segregated hole rich/poor system it is anticipated that the doped holes to be concentrated at the walls \cite{pap} and hence for our case it is expected that Mn$^{4+}$ to be concentrated at the walls. To emphasize the fact that the wall dynamics is modified or suppressed by Al substitution, thermal hysteresis (TH) of ac susceptibility is measured for the Al and Ga substituted samples. Figure 2 shows the imaginary part of ac susceptibility ($\chi$$_1$$^I$) as it is more sensitive to the domain wall dynamics compared to $\chi$$_1$$^R$ which is dominated by the volume response of the domain. Such measurement has also been previously used to probe the change in susceptibility behavior due to modification in the density of spins/holes at the domain walls.\cite{nair} The PM to FM transition is second order in nature, hence, a TH immediately below T$_C$ is ascribed to thermally irreversible domain wall dynamics due to low field irreversible domain wall pinning in the sample.\cite{muk} From inset of Fig. 2, it is observed that a TH is present in $\chi$$_1$$^I$ for the Ga doped sample below T$_C$ (similar to that observed for the parent compound) while, such feature is absent for the Al doped one. This indicates that in the later compound pinning potential of the walls disperses, leading to a reversible behavior in the thermal cycle. This highlights the fact that the wall dynamics is being modified by Al substitution. For 2.5\% Ga compound, the dynamics of walls is similar to the parent as Ga preferentially  replaces the Mn${^3}$$^+$ of the domains. Hence, the above measurements clearly bring out the differences in the intrinsic nature of magnetic state between the Al and Ga substituted samples, which arise from selective substitution of the domain walls and domains respectively.

\subsection{Enhancement and suppression of Griffiths phenomenon by Ga and Al substitution : Role of correlation in disorder}

An important and interesting feature observed in the inset of Fig. 2 is the presence of a peak around 196K for 2.5\% Ga compound.   Such features above T$_C$ is absent for 2.5\% Al sample. Henceforth, we will concentrate on the observed contrasting feature above the T$_C$ for the two substitutions. Recent studies by De Teresa et al.\cite{det} have shown the presence of FM clusters in the PM phase. In 2.5\% Ga compound, the occurrence of the peak around 196K suggests the presence of clusters above T$_C$. This peak remains unchanged with frequency of ac field (Fig. 5). However the nature of the peak changes with the variation of amplitude of ac field (H$_{ac}$), as is observed from the field dependence of $\chi$$_1$$^I$ (Fig 3a). The peak in $\chi$$_1$$^I$ is associated with the magnetic losses and in this case, the observed variation with H$_{ac}$ can be ascribed to the variation of cluster size. The signature of the presence of these clusters is also observed when third order susceptibility ($\chi$$_3$$^R$) is probed; which also shows a peak around 196K (inset Fig 3a). To rule out superparamagnetic (SPM) nature of these clusters, field dependence of this peak in $\chi$$_3$$^R$ is noted. This peak is seen to diverge with the decreasing amplitude of H$_{ac}$ whereas in a SPM it does not diverge as H$_{ac}$$\rightarrow$0.\cite{ash1,nair} Hence, this characteristic along with the absence in any shift in the temperature of this peak with frequency rules out the conventional SPM behavior of the clusters. Another interesting feature of this compound is observed when real part of second order susceptibility ($\chi$$_2$$^R$) is probed. Figure 3b shows the field dependence of temperature response of $\chi$$_2$$^R$ which shows a peak corresponding to peak temperature of 196K found in $\chi$$_1$$^I$ and $\chi$$_3$$^R$. Even order susceptibilities arise in systems in which the inversion symmetry of magnetization (M) with respect applied field is lost i.e. M(H) $\neq$ -M(-H). The signature of peak in $\chi$$_2$$^R$ above T$_C$ ($\approx$ 178K) of this sample clearly indicates the presence of clusters with short-range ordering in the PM region. The local anisotropy within these clusters align the moments in a particular direction which give rise to the symmetry breaking field required for the experimental observation of $\chi$$_2$$^R$. Hence the features in Fig. 3b strongly suggest that the nature of these clusters is FM-type. The peak broadens at higher fields which arises due to the existence of a diluted system of FM clusters resulting in diffuse ordering above T$_C$. Such diffuse ordering is also present in the parent compound indicated by a broad hump in $\chi$$_2$$^R$ around 200K, much above its T$_C$ ($\approx$ 179K) where a sharp peak in $\chi$$_2$$^R$ is observed (inset of Fig 3b). Another evidence of the presence of clusters above T$_C$ comes from the zero field cooled (ZFC) and field cooled (FC) magnetization curves of the sample (Fig 4a). It is observed that the bifurcation between the curves starts much above T$_C$ i.e. from 207K whereas, in typical ferromagnet this bifurcation starts from T$_C$. This feature also demonstrates the presence of FM clusters in the PM phase which results in the bifurcation temperature being higher than T$_C$. Nevertheless, presence of clusters above T$_C$ is not the sufficient condition for GP. It may be noted that non-GP like clustered state above T$_C$ has been observed in La$_{1-x}$Sr$_x$CoO$_3$.\cite{he} The fundamental feature which is typical to GP is the observation of a downward deviation in inverse susceptibility ($\chi$$^{-1}$) from Curie Weiss's (CW) law. Hence, to give a credible explanation to the fact that the clustered state in our system has the characteristic of a GP, inverse $\chi$$_1$$^R$ is plotted as a function of temperature (Fig. 4b). As expected, the curve shows a negative deviation from CW law pointing the fact that GP like features is observed in 2.5\% Ga compound.  The deviation starts from 208K, which is also near to the bifurcation temperature (T) of ZFC and FC curve. In the typical PM regime (T $>$ 208K), the effective PM moment is 4.68 $\mu$$_B$ which aggress with the theoretical value and is also independent of applied field value. Hence, the observed singularity in inverse $\chi$$_1$$^R$ below 208K occur due to the presence of FM clusters in the PM phase. It leads to an enhanced susceptibility compared to the normal PM susceptibility, resulting in the downward deviation in the temperature response of inverse susceptibility. Similar behavior is observed from the inverse dc magnetization data which is qualitatively identical to ac susceptibility data showing a downward deviation from CW behavior (Fig 4c).

This feature in the Ga substituted sample is not unusual since GP like feature is recently observed in the single crystal of the parent composition (La${_{0.9}}$Sr$_{0.1}$MnO${_3}$) by microscopic probes. \cite{dei} Moreover, the observation of a broad hump in $\chi$$_2$$^R$ around 200K  (inset of Fig. 3b) and a downward deviation from the CW law observed in 1/$\chi$$_1$$^R$ vs. T plot (Fig. 4d) confirms the presence of GP like features in the parent compound of the present series. Hence it can be said that addition of Ga enhances the disorder in the parent compound resulting in a more prominent deviation in 1/$\chi$$_1$$^R$ vs. T behavior. This fact is substantiated from the value of susceptibility exponent $\lambda$ deduced from the power-law \cite{sal} of the form $\chi$$^{-1}$ $\propto$ (T/T$_C$ -1)$^{1-\lambda}$, where 0$\leq$$\lambda$$<$1 for GP. It is found that for 2.5\% Ga doped compound, $\lambda$ = 0.23 (Fig. 4e), while that for parent compound $\lambda$ = 0.17 (not shown), clearly indicating that Griffiths singularity in enhanced with Ga substitution. The values of $\lambda$ is similar to those observed in a single crystal of La$_{0.7}$Ba$_{0.3}$MnO$_3$, indicating a weak singularity.\cite{wan} However, for the Al substituted sample GP like features are absent and it follows CW law above T$_C$ (Fig 4f). 

Hence the observed distinct behavior between Ga and Al substituted samples highlights the fact that the presence of correlated disorder is the cause of the observed Griffiths phenomenon. Correlated disorder implies that when disorder is introduced, its effect is distributed in the bulk of the system. The Mn$^{3+}$ ions which constitute significant percentage of the domains, is favorably replaced by Ga, resulting in the effect of disorder being extended to larger length-scale, into the bulk. Hence correlated disorder in the form of Ga results in an enrichment of GP phenomenon in 2.5\% Ga compound. However, this correlation in the effect of disorder is suppressed by preferential replacement of Mn$^{4+}$ by Al, which is in majority at the walls. Such disorder simply modifies the wall and its dynamics. Consequently, no GP phenomenom is observed and we get a clean 1/$\chi$$_1$$^R$ vs. T behavior in the 2.5\% Al compound.

\subsection{Re-entrant spin glass like behavior with Ga and disappearance of the glassy features with Al substitution}

In this section we concentrate on the low temperature phase of the substituted compounds. Even though some glassy dynamics were observed in the OD state of the parent compound, the behavior is not associated to conventional spin/cluster glasses or to any reentrant systems. It is well known that increase of disorder in FM or antiferromagnetic ground state can give rise to frustration leading to special behavior of spin glass (SG) like dynamics in the system.\cite{myd} Such features are observed when Ga is introduced in the system where the temperature of the peak appearing at low temperature in imaginary part of first order ac-$\chi$ ($\chi$$_1$$^I$) is frequency dependent and the peak around T$_c$ does not shift with frequency (Figure 5). The lower temperature peak is shifted downward in temperature with the decrease in frequency, which is a characteristic feature of SG dynamics arising from progressive freezing of spins. The shift of peak temperature with frequency is fitted to power law (which follows from dynamic scaling theory) which is of the form 
\begin{center}
$\tau$=$\tau$$^0$(T/T$_0$-1)$^{-z\nu}$ 
\end{center}
where  $\tau$$^0$ is the microscopic flipping time, T$_0$ is the characteristic temperature of SG freezing, $\nu$ is the critical exponent which describe the growth of spin correlation length and z is the dynamic exponent which describe the slowing down of relaxation (Inset of Fig 5). The fit yielded z$\nu$ = 5.04$\pm$0.35, $\tau$$^0$ = 2.5 *10$^{-8}$ and T$_0$= 99K. The values of $z\nu$ and $\tau$$^0$, which does not match with those reported for SG [generally $z\nu$ lying between 7 to 10 and $\tau$$^0$ of the order of 10$^{-10}$ to 10$^{-13}$ for canonical SG].\cite{sou} Thus it is observed that substitution of Ga in the compound introduces some characteristic features, as observed for SG dynamics but it cannot be associated to any canonical SG systems. 

To provide further evidence about the presence of glassy phase, memory effect similar to Ref [4]  is studied. The experiment is carried out in 4Oe magnetic field with temporary stops (with the magnetic field switched off) at 110K and 95K for a waiting time 7200s. As observed from Fig 6, a decay in magnetization (ageing effect) is noted at the wait temperatures in cooling cycle with the effect being significant at 110K. During the warming cycle a distinct step like feature is seen around 110K indicating that the system remembers its thermal history or the magnetic state reached by the system during the cooling cycle. Such memory effects is widely accepted as a signature of freezing of spins.\cite{vin} Moreover in our system it is expected that long range FM ordering to be present along with the freezing of spins in the low temperature. Such mixed phase is the characteristic of a re-entrant spin glass (RSG).\cite{cha1}

Hence to probe the nature of this glassy phase, temperature response of $\chi$$_2$$^R$ is noted (Fig 7a). The figure, shows a peak around PM to FM transition and a hump at lower temperature. Presence of a significant $\chi$$_2$$^R$ signal in the low temperature region even in absence of DC field supports the presence symmetry breaking field which is an essential feature for the FM ordering. It is to be noted that for a canonical SG, $\chi$$_2$$^R$ is absent. Hence this measurement clearly indicates the presence of mixed phase in the sample where FM ordering coexists with SG freezing. Therefore it can be said that the sample enters a FM phase from a random PM phase and on further lowering of temperature it once again re-enters into a new random phase where FM and SG ordering coexists i.e. a RSG phase. Further, to give a decisive proof about the presence of FM ordering in the low temperature region, Arrott's plot are done on the sample (Fig 7b). Extrapolation of the high field data to H=0 gives the  non zero values of spontaneous magnetization. The curves in the temperature range 85-105K clearly show the presence of spontaneous magnetization which highlights the presence of FM ordering in the low temperature region of the compound. 

The conclusive evidence about the RSG phase comes from the behavior of $\chi$$_3$. This non-linear ac-$\chi$ is used as a probing tool as it is more sensitive to the subtle feature which may remain undetected in temperature response of real part of first order ac-$\chi$. The real part of third order ac-$\chi$ ($\chi$$_3$$^R$) is usually employed  to investigate and differentiate various metamagnetic states exhibited by different compounds.\cite{ash1,sunil} Figure 8 shows the field dependence of $\chi$$_3$$^R$ which shows a peak corresponding to characteristic temperature of SG freezing (T$_0$) at 9Oe. As the amplitude of ac field (H$_{ac}$) is decreased the peak is rounded up unlike in typical SG, where the peak shows a negative divergence.\cite{ash1} Such typical signatures of a non-divergent peak in $\chi$$_3$$^R$ is not uncommon and has also been observed in other reentrant systems.\cite{kun,cha1} Such contrasting behavior at the freezing temperature in the SG and RSG cases arises due to the fact that the transition is approached directly from the PM phase in the former while it is from the FM phase in the latter case. Hence the criticality observed in SG is not observed for RSG due to the existence of FM ordering.

However in 2.5\% Al sample, frequency dependence in $\chi$$_1$$^I$ observed below T$_C$ vanishes in the low temperature region. Moreover no additional peak is observed at low temperature (Fig 9), contrary to that observed in the 2.5\% Ga compound. These features clearly rules out the presence of glassy feature in the low temperature region of this compound. Hence the observation of this section also re-emphasizes the conclusion drawn in the previous sections which highlights the fact that even though impurities are of non-magnetic nature, the disorder effect arising out of these substituents is different. Ga preferentially replacing Mn$^{3+}$ of the domains, enhances both disorder and frustration in the compound which lead to a RSG like behavior. On the other hand, Al favorably replacing the Mn$^{4+}$ of the Mn$^{4+}$ rich walls, introduces uncorrelated disorder resulting in the suppression of glassy dynamics.

\subsection{Similar frequency dependence in $\chi$$_3$$^R$ at low temperature : persistence of orbital domains }

Finally an important feature is observed when temperature response of $\chi$$_3$$^R$ of the entire series is studied (Fig 10). It is observed that $\chi$$_3$$^R$ changes its sign from negative to positive at higher frequencies in the low temperature region of the compound. The changeover in $\chi$$_3$$^R$ is generally observed around T$_C$\cite{sun1} and it arises due to the breaking of spatial magnetic symmetry. Additionally, the frequency dependent crossover of $\chi$$_3$$^R$ implies that this behavior is also dependent upon intra and inter cluster effect.\cite{mat} Such non-linear magnetic response where the intra and inter cluster effects are separated has also been reported for other systems.\cite{miy} Generally the magnetization of each cluster will fluctuate and change the direction with respect to each other in a certain time scale, which is larger than the intra cluster fluctuation. So, at higher frequencies where the probe time is less, the inter cluster fluctuation cannot follow the excitation field and the temperature response of $\chi$$_3$$^R$ reflects the fluctuation inside each individual clusters. 

In the parent compound,  the presence of a low temperature structural transition \cite{ark} introduces an additional degree of freedom (orbital) owing to the change in  occupancy of orbitals by e$_g$ electron, which arises due to the change of lattice constant. This leads to the reformation of domains which results in a change in spin arrangement within the clusters.  The intra cluster ordering, which is formed below T$_C$ is affected by orbital ordering, resulting in a change of spatial magnetic symmetry leading to the frequency dependent crossover of $\chi$$_3$$^R$ [Fig 10(a)]. This typical frequency dependent crossover in $\chi$$_3$$^R$ is also observed in the substituted samples (Fig 10b and 10c). Hence this measurement clearly demonstrates that the intrinsic magnetic configuration of the substituted samples is of the form of orbital domains in analogy to the orbital domain state of the parent compound. Hence it can be said that in all these compounds at low temperature the intra cluster ordering is similar. The observed contrasting magnetic features arise due to a difference in inter cluster interactions which give rise to glassy FM behavior in the parent compound, RSG like features in the 2.5\% Ga compound and absense of glassy characteristic in the 2.5\% Al compound.

\section{Conclusion}  
In summary, we show that, it is possible to selectively introduce disorder either within the domains or at the walls of the orbital domain state in a low doped manganite (La${_{0.9}}$Sr$_{0.1}$MnO${_3}$), without introducing any significant change in the structure. Preferential replacement of Mn$^{3+}$ ions by Ga introduces disorder within the Mn$^{3+}$-rich domains, having its effect extending over the dimension of the domain and is considered to be correlated on similar length-scale. This correlated disorder enhances the GP singularity, whereas, the effect of disorder created by preferential replacement of Mn$^{4+}$ by Al at the hole-rich domain wall is rather restricted. The correlation of such disorder remains confined to the walls; it only modifies the wall dynamics of the system and suppresses the singularity related to GP. In the low temperature region of 2.5\% Ga compound, RSG like features are observed while glassy characteristics are absent in 2.5\% Al compound.   $\chi$$_3$$^R$ measurements are done to highlight the fact that the magnetic state of the entire series are of the form of orbital domains. 

These results are significant since it becomes possible from bulk magnetic measurement to reinforces the nature of self-organization in the low doped manganite and confirm that the hole deficient domains (Mn$^{3+}$ rich) are separated by hole affluent walls (Mn$^{4+}$ rich). It also shows the possibility to selectively create disorder in the self-organized structures in different systems.  Moreover, this study may stimulate investigation on the relation between the nature of disorder and the GP in different systems.

\section{Ackhowledgement} 

We are thankful to P. Chaddah for fruitful discussions. DST, Government of India is acknowledged for funding the commercial VSM. KM acknowledges CSIR, India for financial support.

\begin{figure*}
\centering
\includegraphics[width=8cm]{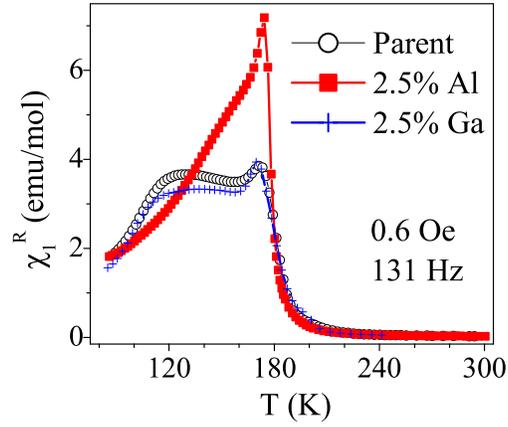}
\caption{(color online) Temperature dependence of the real part of first order ac-susceptibility ($\chi$$_1$$^R$) of the parent, 2.5\% Al and 2.5\% Ga compounds.}
\end{figure*}

\begin{figure*}
\centering
\includegraphics[width=8cm]{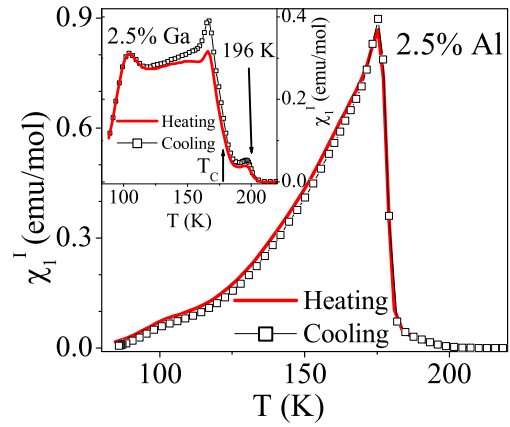}
\caption{(color online) (a) Thermal hysteresis (TH) of imaginary part of first order ac-$\chi$ ($\chi$$_1$$^I$) for 2.5\% Al sample in 1 Oe ac-field of 131 Hz. Inset shows the TH in $\chi$$_1$$^I$ in same measurement condition for 2.5\% Ga sample. The arrows point to the PM-FM transition temperature (T$_C$ $\approx$ 178K) and the temperature of a peak above T$_C$, around 196 K.}
\end{figure*}
\begin{figure*}
\centering
\includegraphics[width=8cm]{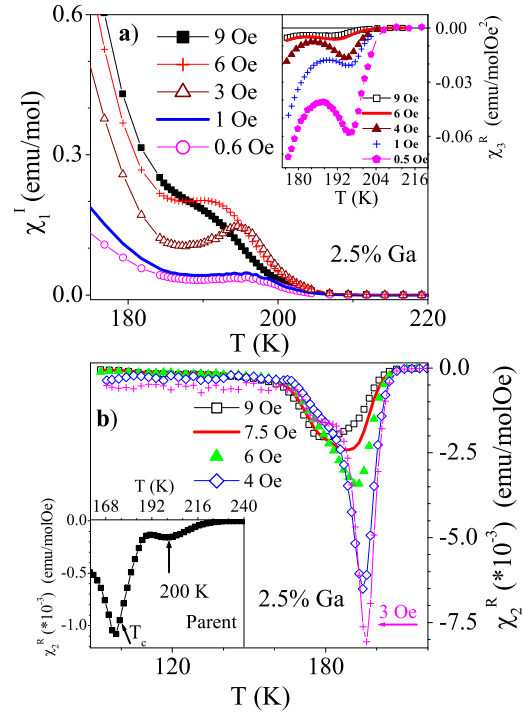}
\caption{(color online) (a) Field dependence of imaginary part of first order AC-$\chi$ ($\chi$$_1$$^I$) of 2.5\% Ga compound at 131 Hz. Inset shows the field dependence of real part of third order ac-$\chi$ ($\chi$$_3$$^R$) for the same sample. b) Field dependence of the real part of second order ac-$\chi$ ($\chi$$_2$$^R$) of 2.5\% Ga sample. Inset shows temperature dependence of the $\chi$$_2$$^R$ of parent compound at 9 Oe ac-field of 131 Hz. The arrows points to the transition temperature (T$_C$ $\approx$ 179K) and the temperature of the hump above T$_C$, around 200 K.}
\end{figure*} 
\begin{figure*}
\centering
\includegraphics[width=8cm]{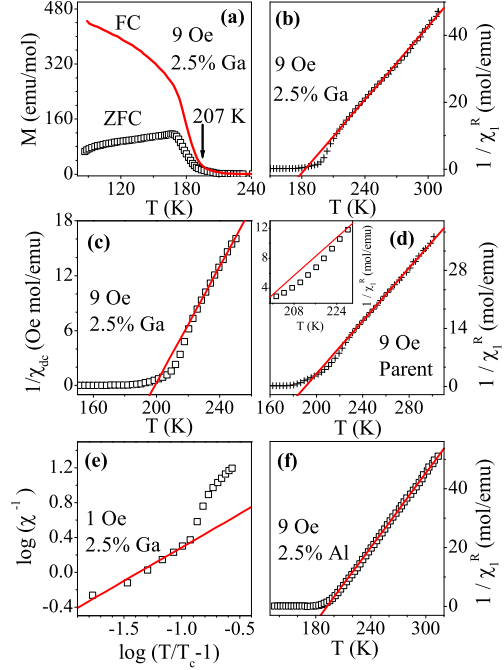}
\caption{(color online)	a) Zero Field Cooled (ZFC) and Field Cooled (FC) dc-magnetization at 9 Oe for the 2.5\% Ga sample. The arrow points the bifurcation temperature, around 207 K, between ZFC and FC curves. b) Temperature dependence of the inverse of real part of first order ac-$\chi$ ($\chi$$_1$$^R$) at 9 Oe ac-field for 2.5\% Ga sample.  c) Temperature dependence of the inverse dc-susceptibility at 9Oe dc-field for 2.5\% Ga sample. d) Temperature dependence of the inverse of $\chi$$_1$$^R$ at 9Oe ac-field for the parent compound. Inset shows the magnified view around the inflexion. e)Log-log plot of the inverse of $\chi$$_1$$^R$ as a function of reduced temperature yielding susceptibility exponent $\lambda$=0.23 for 2.5\% Ga sample. f) Temperature dependence of the inverse of $\chi$$_1$$^R$ at 9Oe ac-field for the 2.5\% Al doped sample. Solid lines in (b), (c), (d) and (f) are Curie Weiss fit to the high-T behavior.}  
\end{figure*}	
\begin{figure*}
	\centering
		\includegraphics[width=8cm]{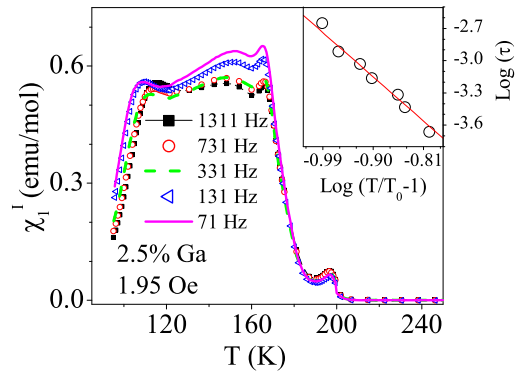}
	\caption{(color online) Frequency dependence of imaginary part of AC-$\chi$ ($\chi$$_1$$^I$) of 2.5\% Ga sample.  Inset : Dynamical scaling fit of peak temperature with reduced temperature.}
\end{figure*}
\begin{figure*}
	\centering
		\includegraphics[width=8cm]{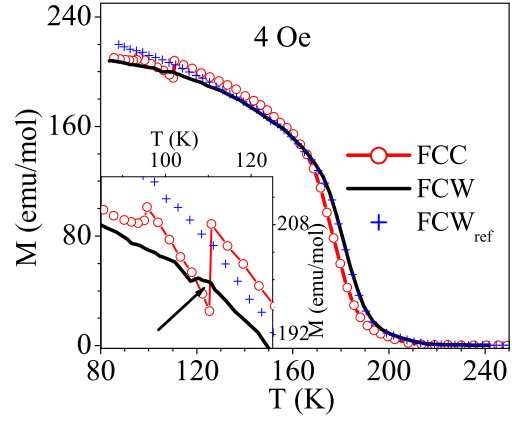}
	\caption{(color online) M-T curves during field cooling of 2.5\% Ga sample. The field was switched off at two temperatures (110K and 95K) for a waiting time of 7200 s. The M-T curve in warming mode and normal FCW curve as FCW$_{ref}$ is also shown. Inset shows the expanded view of the memory effect. The arrow points to step like feature observed in the warming cycle (FCW).}
\end{figure*}
\begin{figure*}
	\centering
		\includegraphics[width=8cm]{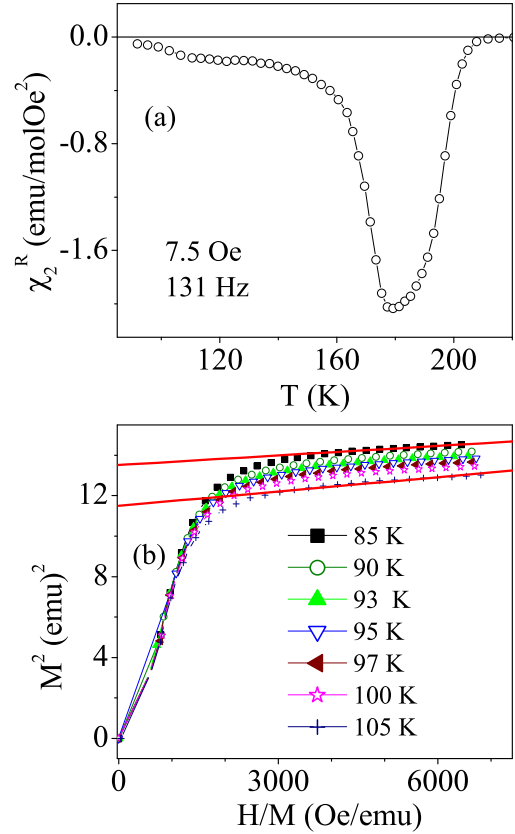}
	\caption{(color online) (a) Temperature response of real part of second order susceptibility ($\chi$$_2$$^R$) of 2.5\% Ga sample. (b) Arrott's plot in the temperature range 85-105K for the same. Solid lines are linear fit to the M$^2$ vs H/M curve indicating the presense of spontaneous magnetization.}
\end{figure*}
\begin{figure*}
	\centering
		\includegraphics[width=10cm]{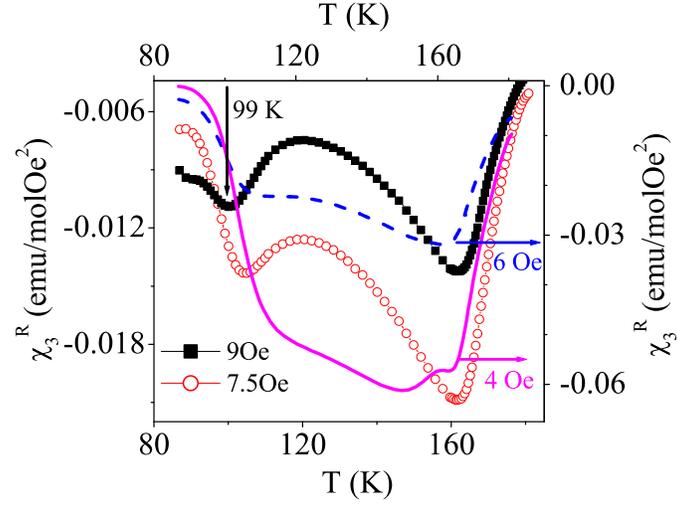}
	\caption{(color online) Temperature response of real part of third order susceptibility ($\chi$$_3$$^R$) of 2.5\% Ga sample. The black arrow points to the peak temperature of 99K at 9Oe indicating spin-glass like freezing.}
\end{figure*}
\begin{figure*}
	\centering
		\includegraphics[width=8cm]{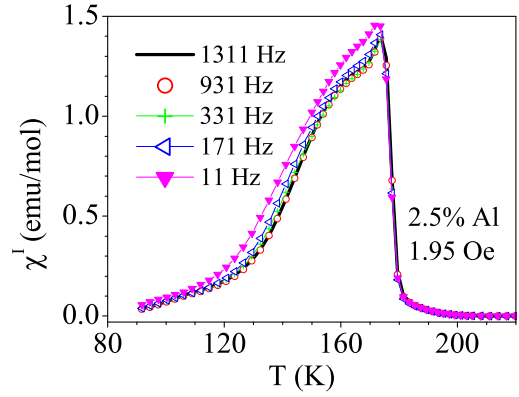}
	\caption{(color online) Temperature response of frequency dependence of imaginary part of ac-$\chi$ ($\chi$$_1$$^I$) of 2.5\% Al sample. The frequency dependence collapses in low temperature region.}
\end{figure*}
\begin{figure*}
	\centering
		\includegraphics[width=8cm]{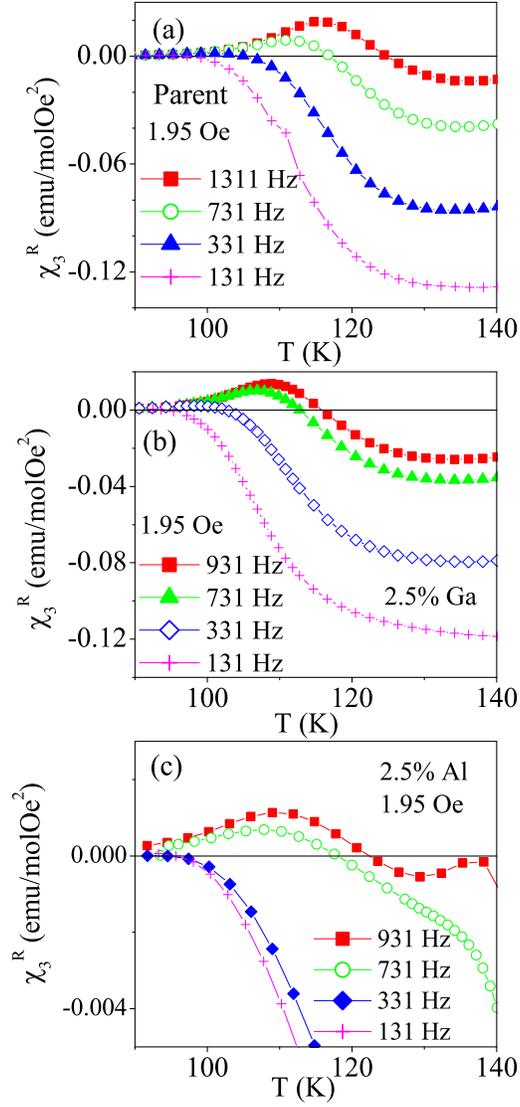}
	\caption{(color online) Temperature reponse of frequency dependence of real part of third order of ac-$\chi$ ($\chi$$_3$$^R$) for the a) parent b) 2.5\% Ga and c) 2.5\% Al compounds. All of them show crossover of $\chi$$_3$$^R$ in higher frequencies.}
\end{figure*}
\end{document}